\documentclass[aps,prb,twocolumn]{revtex4}
\usepackage{amsmath,amssymb,mathrsfs}
\usepackage{color}
\usepackage{psfig}
\begin{document}

\title{Quantum theory of the electromagnetic response of metal
nanofilms}

\author{George Y. Panasyuk~\footnote{Formerly, at the Department of
    Bioengineering, University of Pennsylvania, Philadelphia, PA
    19104}}
\affiliation{Propulsion Directorate, Air Force Research Laboratory,
  Wright-Patterson Air Force Base, OH 45433} 
\author{John C. Schotland} 
\affiliation{Department of Mathematics, University of Michigan, Ann Arbor, MI 48109} 
\author{Vadim A. Markel} 
\affiliation{Departments of Radiology, Bioengineering, and
  Graduate Group in Applied Mathematics and Computational Science,
  University of Pennsylvania, Philadelphia, PA 19104}

\begin{abstract}
  We develop a quantum theory of electron confinement in metal
  nanofilms. The theory is used to compute the nonlinear response of
  the film to a static or low-frequency external electric field and to
  investigate the role of boundary conditions imposed on the metal
  surface. We find that the sign and magnitude of the nonlinear
  polarizability depends dramatically on the type of boundary
  condition used.
\end{abstract}

\maketitle

\section{Introduction}
\label{sec:intro}

The recent explosion of growth in the field of
nanoplasmonics~\cite{maier_03_1,prodan_03_1,stockman_08_1,stockman_08_2,brongersma_10_1}
have been a joint success of theory and experiment. Yet, in certain
respects, theory is lagging behind. One profound theoretical question,
which has not received adequate attention so far, is related to the
applicability of macroscopic theories. That is, the theory of
plasmonic systems is almost exclusively based on the macroscopic
Maxwell's equations, even though the samples involved are, in some
cases, only a few nanometers in size. The problem is compounded by the
fact that plasmonic applications utilize highly conductive noble
metals. In this case, the mean free path of the conduction electrons,
which can be significantly larger than interatomic distances, becomes
of primary importance. Several different approaches to accounting for
the finite-size and quantum effects in nanoparticles have been
proposed~\cite{halas_11_1}. However, finite-size-dependent
electromagnetic nonlinearities have received relatively little
attention. In this paper, we theoretically investigate finite-size
effects in metallic nanofilms with an emphasis on electromagnetic
nonlinearity and on the role of boundary conditions applied at the
nanofilm surface.

A qualitative understanding of finite-size effects in plasmonic
systems can be gained from purely classical arguments. For example, it
has been demonstrated that the electromagnetic response of a strongly
coupled plasmonic system is dramatically altered by the effects of
nonlocality when the smallest geometrical feature of the system (e.g.,
an interparticle gap) becomes comparable to a certain phenomenological
length scale, which characterizes the intrinsic nonlocality of
metal~\cite{dasgupta_81_1,fuchs_87_1,ruppin_92_1}.  A qualitative
theory of optical nonlinearity can also be derived from purely
classical arguments~\cite{panasyuk_08_1}. 

However, a quantitatively accurate theory must be quantum mechanical.
Unfortunately, a fully self-consistent and mathematically-tractable
quantum model of plasmonic systems is difficult to formulate. In
Refs.~\onlinecite{hache_86_1,rautian_97_1}, a metal nanosphere was
approximated by a degenerate electron gas confined in a spherical,
infinitely-high potential well. The conduction electrons were assumed
to be driven by a time-harmonic and spatially-uniform electric field.
This driving field is internal to the nanoparticle; relating it to the
external (applied) field constitutes a separate
problem~\cite{drachev_04_2}. The model of
Refs.~\onlinecite{hache_86_1,rautian_97_1} has attracted considerable
attention in the optics community. In particular, it predicts
correctly the size dependence of the Drude relaxation constant.
Although the model is mathematically tractable, it yields an
expression for the third-order nonlinear polarizability, which
involves an eight-fold series. This expression can be evaluated only
approximately~\cite{rautian_97_1}. Recently, we have reduced
analytically this expression (without resorting to any additional
approximations) to a five-fold series, which is amenable to direct
numerical evaluation~\cite{govyadinov_11_1}. We then have used this
result to study the size- and frequency dependence of the third-order
nonlinear polarizability.

Still, formulation of the model of
Refs.~\onlinecite{hache_86_1,rautian_97_1} involves two important
assumptions, which are difficult to avoid and whose effect is
difficult to predict. First, the model assumes a spatially-uniform
electric field inside the nanoparticle. In reality, this field is not
spatially-uniform because the induced charge density is different from
the infinitely-thin surface density of the Lorenz-Lorentz theory,
which was used in Refs.~\onlinecite{rautian_97_1,drachev_04_2} and
also by us~\cite{govyadinov_11_1} to obtain a relation between the
internal and the applied fields. A relation of this kind is essential
for deriving any experimentally-measurable quantity. Unfortunately, a
more fundamental approach to relating the internal and applied fields
does not readily present itself, at least, not within the formalism of
Refs.~\onlinecite{hache_86_1,rautian_97_1}. Second, the imposition of
an infinite potential barrier at the nanoparticle surface has not been
justified from first principles, even though this boundary condition
is frequently employed and can be defended by noting that a gas of
noninteracting electrons can not be stably bound by the Coulomb
potential of a spatially-uniform, positively-charged background. In
other words, the potential barrier makes up for the neglect of the
discrete nature of positively-charged ions.

The two assumptions described above can be, in principle, avoided by
using the density-functional theory (DFT). Within DFT, the
exchange-correlation potential renders the system stable even if the
ionic lattice is replaced by jellium. Time dependent DFT (TDDFT) and
the jellium model have been successfully used to compute the linear
response of nanoparticles~\cite{ekardt_84_1,lerme_99_1}. TDDFT and the
random-phase approximation (RPA) have been used without resorting to
jellium model (that is, with the full account of crystalline structure
of metal) to study the effects of surface molecular adsorption on the
dielectric losses in metal~\cite{zhu_08_1,gavrilenko_10_1}. Size- and
shape-dependence of the imaginary part of the dielectric function of
Ag nanoparticles has been also studied
experimentally~\cite{drachev_08_1,qiu_08_1}.

However, it is not clear whether the same DFT-based approaches are
appropriate for obtaining nonlinear corrections to the polarizability.
Indeed, the exchange-correlation potential is typically constructed
for an infinite and spatially-uniform system and is not expected to be
accurate near boundaries, where the electron density changes on small
spatial scales. Yet, it can be argued that the nonlinear response of a
nanoparticle is strongly influenced by the electron density near the
boundary~\cite{panasyuk_08_1}. Further, if the jellium model is used,
as is the case in this work, the binding potential of the
positively-charged jellium (even with the account of
exchange-correlation potential) is not as strong as that of discrete
ionic lattice. In the latter case, the Coulomb potential approaches
negative infinity in the vicinity of the ion cores.  Therefore, it is
not evident whether the correct model should incorporate an additional
potential barrier at the surface of the nanoparticle to account phenomenologically for the
reduced binding power of the jellium.  Considering the above
uncertainty, and the fact that the infinite potential barrier of
Ref.~\onlinecite{rautian_97_1,drachev_04_2} has been widely used in
the optics literature, it seems desirable to investigate the influence
of the boundary conditions used on the obtained nonlinear corrections.

To address the problem formulated above, we have performed DFT
calculations for the simplest plasmonic system -- a thin metallic film
to which a static or low-frequency perpendicularly-polarized electric
field is applied. We have investigated the influence of three
different types of boundary conditions, including the rigid (infinite
potential barrier) boundary condition, the Bardeen's boundary
condition (in which the potential barrier is displaced from the metal
surface) and free boundaries. We compute nonlinear corrections to the
film polarizability by two different approaches -- first, by the use
of a perturbation theory, which is described in detail below and,
second, by direct numerical application of the DFT (for control). For
relatively strong applied fields, when the perturbation theory is not
applicable, the nonlinear polarization is computed directly by DFT.

The paper is organized as follows. In Sec.~\ref{sec:model} we describe
the physical model in more detail. In Sec.~\ref{sec:DFT}, we write the
basic equations of the DFT and specialize them to the thin film
problem. The perturbation theory is described in Sec.~\ref{sec:pert}.
Numerical results are reported in Sec.~\ref{sec:res}. Finally,
Sec.~\ref{sec:summ} contains a summary and a discussion.

\section{The physical model}
\label{sec:model}

Throughout this paper we consider the following model system. A thin
metallic film of width $h$ is taken to occupy the spatial region $-L/2
<x,y <L/2$, $-h/2 < z < h/2$, where $L \gg h$. The latter condition
allows for the neglect of effects due to the edges of the slab.
Accordingly, we assume that all physical quantities depend only on the
variable $z$, so that the system is effectively one-dimensional. A
static electric field of strength ${\mathscr E}$ is applied to the
slab in the $z$-direction. We emphasize that this is the external
(applied) field; not the internal field of
Refs.~\onlinecite{hache_86_1,rautian_97_1}.

We use the stabilized jellium model~\cite{shore,perdew,liebsch} and
the Gunnarsson-Lundqvist local-density approximation for the
exchange-correlation potential~\cite{gunnarsson_76_1} (defined
precisely below). At or near the film surface, we apply three
different types of boundary conditions.  In our first set of
calculations, we apply the infinite potential wall condition of
Refs.~\onlinecite{hache_86_1,rautian_97_1} at the physical boundary of
metal. We denote this type of boundary condition by ``R'' (in memory
of S.G.Rautian~\cite{rautian_97_1}).  In a second set of calculations,
employing what we refer to as the ``B''-type boundary condition, we
use the classical Bardeen model~\cite{bardeen,pitarke_eguiluz,han} in
which the potential wall is displaced from the physical boundary by
the distance
\begin{equation}
\label{Delta_B}
\Delta_{\rm B} = \frac{3\pi}{8k_{\rm F}} \ , 
\end{equation}
where $k_{\rm F}$ is the Fermi wave number. The displacement
$\Delta_{\rm B}$ can be expressed in terms of the characteristic
length scale of the problem, $\ell$, where $\ell^3$ is the specific
volume per conduction electron.  Using the expressions $\hbar k_{\rm
  F} = \sqrt{2 m_e E_{\rm F}}$ and $E_{\rm F} =
(3\pi^2)^{2/3}\hbar^2/2m_e\ell^2$, with $E_{\rm F}$ being the Fermi
energy and $m_e$ the electron mass, we find that $\Delta_{\rm B} =
(9\pi)^{1/3} 8^{-1} \ell \approx 0.38\ell$.  Finally, in the third
set of calculations, we do not use any additional potential barriers
at or close to the metal surface. We denote this last boundary
condition as type-``F''.

In all these cases, we compute the electromagnetic response
(polarization). The quantity of interest is the induced dipole moment
per unit area of the slab ${\mathscr P}$, which is related to the
charge density of the conduction electrons $-\rho_e(z)$, by
\begin{equation}
\label{dipole}
{\mathscr P} = - \int z \rho_e(z) dz \ .
\end{equation}
\noindent
The quantity $\rho_e(z)\geq 0$ has been computed using standard DFT
theory using the three different types of boundary conditions, which
are described above. We have investigated the dependence of ${\mathscr
  P}$ on the strength of the applied field and on the film width. When
nonlinear corrections to ${\mathscr P}$ are computed, the
characteristic scale for the applied field strength is the atomic
field
\begin{equation}
\label{E_at}
{\mathscr E}_{\rm at} = \frac{e}{\ell^2} \ .
\end{equation}
The surface density of the dipole moment can then be conveniently
normalized by the quantity 
\begin{equation}
{\mathscr P}_{\rm at} = \frac{{\mathscr E}_{\rm at}h}{4\pi} \ .
\end{equation}
For a relatively small applied field, the induced polarization can be
expanded in powers of ${\mathscr E}/{\mathscr E}_{\rm at}$ according to
\begin{equation}
\label{exp_1}
{\mathscr P} = \frac{h{\mathscr E}}{4\pi}\left[ \alpha_1  + 
\alpha_3 \left(\frac{{\mathscr E}}{{\mathscr E}_{\rm at}}\right)^2 + \alpha_5
\left( \frac{{\mathscr E}}{{\mathscr E}_{\rm at}}\right)^4 + \cdots \right] \ .
\end{equation}
Note that the coefficients $\alpha_2$, $\alpha_4$, etc., are identically zero in
the above expansion due to the slab symmetry.

The slab width can be parametrized by the number of atomic layers,
$M$. Many metals of interest in plasmonics (including silver) have an
fcc lattice structure with four conduction electrons per unit cell. In
this case, the slab width is $h = M a$, where the lattice step $a$ is
related to $\ell$ by $a = 4^{1/3}\ell$. The slab then occupies the
region $-M a/2 < z < M a/2$. In silver, $\ell \approx 0.26{\rm nm}$ and
$a\approx 0.41{\rm nm}$.

\section{Basic equations}
\label{sec:DFT}

The starting point in DFT is the
eigenproblem~\cite{kohn_65_1,martin_book_04}:
\begin{equation}
\label{ansatz}
\left [-\frac{\hbar^2}{2m_e}\nabla^2 + U_{\rm eff}(z)\right ]\psi_\mu({\bf r}) = 
E_\mu \psi_\mu({\bf r}) \ ,
\end{equation}
\noindent
where the index $\mu$ labels energy eigenstates, $m_e$ is the electron
mass and
\begin{equation}
\label{U_eff}
U_{\rm eff}(z) = U_{\rm H}(z) + U_{\rm xc}(z) + V(z)
\end{equation}
\noindent
is the effective potential consisting of the Hartree term $U_{\rm
  H}(z)$, the exchange-correlation potential $U_{\rm xc}(z)$, and of
the interaction potential 
\begin{equation}
\label{V_z}
V(z) = -ez{\mathscr E} \ ,
\end{equation}
\noindent
which describes interaction of the electrons with the applied field.

We adopt the stabilized jellium model~\cite{shore,perdew,liebsch},
according to which the positively-charged ions form a
spatially-uniform charge density $\rho_i(z)$, such that $\rho_i(z) =
e/\ell^3$ if $\vert z \vert \leq h/2$ and $\rho_i(z) = 0$ otherwise.
In the stabilized jellium model, a constant potential (in our case,
negative-valued) is added to $U_{\rm eff}(z)$ inside the spatial
region occupied by jellium.  This constant potential is chosen so as
to make the metal mechanically stable at its observed valence electron
density. As we found, the difference in the results obtained using the
usual and the stabilized jellium models is insignificant (it is absent
for the R-type boundary condition). On the other hand, the use of the
stabilized jellium model removes certain anomalies of the standard
jellium model, such as negative surface energy~\cite{perdew}. Also, in
the case of a silver film, we have used the stabilized jellium to
compute the work function, which we find to be $3.8{\rm eV}$ for the
thickest film considered (with $M=32$ atomic layers). This result can
be compared to the experimental measurements in
silver~\cite{chelvayohan} [$4.2{\rm eV}$ for the (100) face]. We have
obtained a somewhat less accurate prediction without jellium
stabilization ($3.5{\rm eV}$). Therefore, the stabilized jellium model
is used in all calculations reported below.

The Hartree potential is given by
\begin{equation}
\label{hartree}
U_{\rm H}(z) = 2 \pi e \int_{-\infty}^{\infty} 
[\rho_i(z^\prime) - \rho_e(z^\prime)] \vert z - z^\prime \vert dz^\prime \ ,
\end{equation}
\noindent
where $-\rho_e(z)$ is the density of charge associated with the
conduction electrons. Both densities are normalized by the condition
$\int \rho_i(z) dz = \int \rho_e(z)dz = eh/\ell^3$.

The Gunnarsson-Lundqvist local-density approximation for the
exchange-correlation potential, which we use in this work, is of the
following form:
\begin{equation}
\label{U_xc_def}
U_{\rm xc}(z) = - \left[C_{\rm x} \frac{e^2}{R} + C_{\rm
    c}\frac{e^2}{a_B}\ln\left(1 + A \frac{a_B}{R} \right) \right] \ .
\end{equation}
\noindent
Here $R = (3e/4\pi \rho_e)^{1/3}$, $a_B=\hbar^2/m_e e^2$ is the Bohr
radius, and the dimensionless constants are $C_{\rm x}=0.61$, $C_{\rm
  c} = 0.033$, and $A = 11.4$.

Assuming periodic boundary condition at $x=\pm L/2$ and $y=\pm L/2$,
we can write the eigenfunctions of (\ref{ansatz}) in the form
\begin{equation}
\label{xy_dep}
\psi_{lmn}({\bf r}) = 
\frac{1}{L}{\rm exp}\left [ i\frac{2\pi}{L}(l x + m y)\right ]
\phi_n(z) \ ,
\end{equation}
\noindent
where $\phi_n(z)$ satisfies the equation
\begin{equation}
\label{ansatz_z}
-\frac{\hbar^2}{2m_e}\phi_n^{\prime\prime}(z) + U_{\rm eff}(z)\phi_n(z) = 
\epsilon_n\phi_n(z) \ .
\end{equation}
\noindent
In Eq.~(\ref{xy_dep}), we have replaced the index $\mu$ by the triplet
of quantum numbers $(l,m,n)$. In what follows, we view $\mu$ as a
composite index. For example, summation over $\mu$ entails summation
over the three quantum numbers $l$, $m$ and $n$. The energy
eigenvalues are given by
\begin{equation}
\label{E_lmn}
E_{lmn} = (2 \pi \hbar)^2 \frac{l^2 + m^2}{2m_e L^2} + \epsilon_n \ .  
\end{equation}
The ground-state electron density is given by the expression
\begin{align}
\rho_e(z) & = 2e \sum_\mu \Theta(E_{\rm F} - E_\mu) \vert
\psi_\mu({\bf r}) \vert^2 \nonumber \\
          & = e \sum_{n=1}^{n_{\rm F}} W_n \left \vert
            \phi_n(z) \right\vert^2 \ .  
\label{ground}
\end{align}
\noindent
Here the factor of $2$ in the first equality in (\ref{ground})
accounts for the electron spin, $E_{\rm F}$ is the Fermi energy,
$n_{\rm F}$ is the maximum quantum number $n$ for which there exist
energy levels $E_{lmn}$ below the Fermi surface, and $W_n$ is a
statistical weight given by
\begin{equation}
\label{Wn}
W_n = \frac{m_e(E_{\rm F} - \epsilon_n)}{\pi \hbar^2} \ .
\end{equation}
\noindent
Note that the Fermi energy $E_{\rm F}$ for the model considered in
this paper is close to the macroscopic limit of $(\hbar^2/2m_e)(
3\pi^2 / \ell^3 )^{2/3}$. In all calculations shown in the paper, we
have computed $E_{\rm F}$ and the related quantity $n_{\rm F}$ by
ordering all energy levels $E_{lmn}$. It should be also emphasized
that the quantity $n_{\rm F}$ is determined with the account of all
degeneracies, including those due to the spin, so that the second
equality in (\ref{ground}) does not contain the factor of two. 

Once $\rho_e(z)$ is found, the dipole moment per unit area of the slab
${\mathscr P}$ can be computed according to (\ref{dipole}).

\section{Perturbation theory}
\label{sec:pert}

A perturbative solution to the eigenproblem (\ref{ansatz_z}) is obtained by
expanding the quantities $\phi_n(z)$, $\epsilon_n$, $W_n$, $U_{\rm
  eff}(z)$, $\rho_e(z)$ into power series in the variable $x =
{\mathscr E}/{\mathscr E}_{\rm at}$.  For example, we write
\begin{equation}
\label{E_expan}
\phi_n(z) = \sum_{s=0}^\infty \phi_n^{(s)}(z) x^s \ .
\end{equation}
\noindent
Upon substitution of these expansions in the original eigenproblem
(\ref{ansatz_z}), we obtain the following relations (to third order in
$x$):
\begin{subequations}
\label{E_expan1}
\begin{align}
\label{E_expan1_0}
& s = 0: \nonumber \\
& H_0\phi_n^{(0)} =  \epsilon_n^{(0)}\phi_n^{(0)} \ , \\
\label{E_expan1_1}
& s = 1: \nonumber \\
& H_0 \phi_n^{(1)} + U_{\rm eff}^{(1)}\phi_n^{(0)} =
\epsilon_n^{(0)}\phi_n^{(1)} + \epsilon_n^{(1)}\phi_n^{(0)} \ , \\
\label{E_expan1_2}
& s = 2: \nonumber \\
& H_0\phi_n^{(2)} + U_{\rm eff}^{(1)}\phi_n^{(1)} + U_{\rm eff}^{(2)}\phi_n^{(0)}
= \epsilon_n^{(0)}\phi_n^{(2)} + \epsilon_n^{(1)}\phi_n^{(1)} \nonumber \\
& \hspace*{2cm} + \epsilon_n^{(2)}\phi_n^{(0)} \ , \\ 
\label{E_expan1_3}
& s = 3: \nonumber \\
&  H_0 \phi_n^{(3)} + U_{\rm eff}^{(1)}\phi_n^{(2)} + U_{\rm eff}^{(2)}\phi_n^{(1)}
+ U_{\rm eff}^{(3)}\phi_n^{(0)} = \epsilon_n^{(0)}\phi_n^{(3)}  \nonumber \\
& \hspace*{2cm} + \epsilon_n^{(1)}\phi_n^{(2)} +
\epsilon_n^{(2)}\phi_n^{(1)} + \epsilon_n^{(3)}\phi_n^{(0)} \ .
\end{align}
\end{subequations}
\noindent
Here
\begin{equation}
\label{H_0}
H_0 = -\frac{\hbar^2}{2m_e}\partial_z^2 + U_H^{(0)} + U_{\rm xc}^{(0)}
\end{equation}
\noindent
is the unperturbed Hamiltonian and 
\begin{align}
U_{\rm eff}^{(s)}(z) & = 2\pi e \int_{-\infty}^{\infty} \vert z - \xi
\vert \rho_e^{(s)}(\xi) d \xi \nonumber \\
& + U_{\rm xc}^\prime (z) \rho_e^{(s)}(z)
+ V^{(s)}(z) \ ,
\label{V_expan}
\end{align}
\noindent
where
\begin{subequations}
\label{DeltaV}
\begin{align}
\label{DeltaV_0}
& V^{(0)}(z) = 0 \ , \\
\label{DeltaV_1}
& V^{(1)}(z) = -ez{\mathscr E}_{\rm at} \ , \\
\label{DeltaV_2}
& V^{(2)}(z) = \frac{1}{2}U_{\rm xc}^{\prime\prime}(z) [\rho_e^{(1)}(z)]^2 \ , \\
\label{DeltaV_3}
& V^{(3)}(z) = U_{\rm xc}^{\prime\prime}(z) \rho_e^{(1)}(z) \rho_e^{(2)}(z)
+ \frac{1}{6}U_{\rm xc}^{\prime\prime\prime}[\rho_e^{(1)}(z)]^3 .
\end{align}
\end{subequations}
\noindent
The operators $U_{\rm xc}^{\prime}$, $U_{\rm xc}^{\prime\prime}$, and
$U_{\rm xc}^{\prime\prime\prime}$ are the first, second, and third
functional derivatives of the exchange-correlation potential $U_{\rm
  xc}$ with respect to $\rho_e$ evaluated at $\rho_e = \rho_e^{(0)}$.

In deriving (\ref{V_expan}),(\ref{DeltaV}), we have taken into account
the implicit dependence of $U_{\rm eff}$ on the expansion parameter
$x$, which stems from the dependence of $\rho_e$ on $x$. The first
four coefficients in the expansion of $\rho_e$ can be obtained using
equations (\ref{ground}) and (\ref{Wn}), which results in
\begin{subequations}
\label{n_s}
\begin{align}
\label{n_1}
\rho_e^{(1)} = & e \sum_{n=1}^{n_{\rm F}} \left[ W_n^{(1)} \left(
    \phi_n^{(0)} \right)^2 +
2 W_n^{(0)} \phi_n^{(0)} \phi_n^{(1)} \right] \ , \\ 
\label{n_2}
\rho_e^{(2)} = & e \sum_{n=1}^{n_{\rm F}} \left[ W_n^{(2)} \left(\phi_n^{(0)}\right)^2 +
 2 W_n^{(0)} \phi_n^{(0)} \phi_n^{(2)} \right. \nonumber \\ 
 & \left. + W_n^{(0)} \left(\phi_n^{(1)}\right)^2
\right] \ , \\ 
\label{n_3}
\rho_e^{(3)} = & e \sum_{n=1}^{n_{\rm F}} \left[
  2W_n^{(0)} \left(\phi_n^{(0)}\phi_n^{(3)} + \phi_n^{(1)}
    \phi_n^{(2)} \right) + 
\right. \nonumber \\
& \left. 2W_n^{(2)}\phi_n^{(0)}\phi_n^{(1)} +  W_n^{(3)}\left(\phi_n^{(0)} \right)^2 \right] \ .
\end{align}
\end{subequations}
\noindent
Here 
\begin{equation}
\label{W_expan}
W^{(s)}_n = w
\left [ \sum_{k=1}^{n_{\rm F}}\epsilon_k^{(s)} -
  n_{\rm F}\epsilon_n^{(s)}\right ]
\end{equation}
\noindent
are the coefficients in the expansion of the statistical weights
$W_n$ and
\begin{equation}
w = \frac{m_e}{\pi n_{\rm F}\hbar^2} \ .
\end{equation}
We note that the integer number $n_{\rm F}$, which was defined (after
Eq.~\ref{ground}) as the maximum quantum number $n$ for which there
exist energy levels $E_{lmn}$ below the Fermi surface, is independent
of the applied field ${\mathscr E}$ for ${\mathscr E}\lesssim
10 {\mathscr E}_{\rm at}$. Therefore, in the perturbation theory
developed here, it is sufficient to compute $n_{\rm F}$ at zero
applied field.

The procedure of constructing the perturbation series for the dipole
moment density, ${\mathscr P}$, starts from solving (\ref{E_expan1_0})
numerically. The eigenvalues $\epsilon_n^{(0)}$ are ordered so that
$\epsilon_n < \epsilon_{n+1}$ for $n = 1, 2, ...$. It
follows from symmetry considerations that
\begin{subequations}
\label{parity_0}
\begin{align}
\label{parity_0phi}
& \phi_n^{(0)}(-z) = (-1)^{(n-1)}\phi_n^{(0)}(z) \ , \\
\label{parity_0n}
& \rho_e^{(0)}(-z) = \rho_e^{(0)}(z) \ .
\end{align}
\end{subequations}
\noindent
These relations have been strictly enforced in the numerical
procedures we have implemented.
 
We seek the solutions to (\ref{E_expan1}) in the following form:
\begin{equation}
\label{phi_s}
\vert \phi_n^{(s)} \rangle = \sum_{k}C_{nk}^{(s)}\vert \phi_k^{(0)} \rangle \ .
\end{equation}
\noindent
Here the unperturbed basis functions $\vert \phi_k^{(0)} \rangle$ (and
the energy levels $\epsilon_n^{(0)}$) are assumed to be known; they
are determined by solving (\ref{E_expan1_0}) numerically.  In the
zeroth order ($s=0$), we have trivially $C_{nk}^{(0)}=\delta_{nk}$. In
higher expansion orders, we determine $C_{nk}^{(s)}$ by substituting
(\ref{phi_s}) into (\ref{E_expan1}). For $s=1,2,3$, this results in
the following three equations:
\begin{subequations}
\label{phi}
\begin{align}
\label{phi_1}
\vert \phi_n^{(1)} \rangle & = \sum_{k \neq n} 
\frac{ \vert \phi_k^{(0)} \rangle}{\epsilon_n^{(0)} -  \epsilon_k^{(0)}} 
\langle \phi_k^{(0)} \vert U_{\rm eff}^{(1)} \vert \phi_n^{(0)} \rangle  \ , \\ 
\label{phi_2}
\vert \phi_n^{(2)} \rangle & = \sum_{k \neq n}
\frac{ \vert \phi_k^{(0)} \rangle}{\epsilon_n^{(0)} -  \epsilon_k^{(0)}} 
\left[
  \langle \phi_k^{(0)} \vert U_{\rm eff}^{(1)} \vert \phi_n^{(1)} \rangle
\right. \nonumber \\ 
& \left. + \langle \phi_k^{(0)} \vert U_{\rm eff}^{(2)} \vert \phi_n^{(0)} \rangle 
\right] 
 - \frac{1}{2}
\vert \phi_n^{(0)} \rangle \langle \phi_n^{(1)} \vert \phi_n^{(1)}\rangle \ , \\ 
\label{phi_3}
\vert \phi_n^{(3)} \rangle & = \sum_{k \neq n}
\frac{\vert \phi_k^{(0)} \rangle}{\epsilon_n^{(0)} - \epsilon_k^{(0)}}
\left[ 
    \langle \phi_k^{(0)} \vert U_{\rm eff}^{(1)} \vert \phi_n^{(2)} \rangle
\right. \nonumber \\
& 
  + \langle \phi_k^{(0)} \vert U_{\rm eff}^{(2)} \vert \phi_n^{(1)} \rangle 
  + \langle \phi_k^{(0)} \vert U_{\rm eff}^{(3)} \vert \phi_n^{(0)} \rangle
  - \langle \phi_k^{(0)} \vert \phi_n^{(1)} \rangle
\nonumber \\
& \times \left. \left( \langle \phi_k^{(0)} \vert U_{\rm eff}^{(2)} \vert \phi_n^{(0)} \rangle +
\langle \phi^{(0)}_n\vert U_{\rm eff}^{(1)} \vert \phi_n^{(1)} \rangle
\right)
\right ] 
 \ .
\end{align}
\end{subequations}
\noindent
Regarding the set of equations (\ref{phi}), we note the following.
First, the vectors $\vert \phi_k^{(s)} \rangle$ must be computed
recursively starting with $s=1$. Thus, we compute $\vert \phi_n^{(1)}
\rangle$ according to (\ref{phi_1}). The result is substituted into
(\ref{phi_2}), which allows one to compute $\vert \phi_n^{(2)}
\rangle$, and so forth.  However, unlike in the ordinary perturbation
theory, the operators $U_{\rm eff}^{(s)}$ in the right-hand sides of
(\ref{phi}) are still unknown and must be determined separately.
Indeed, $U_{\rm eff}^{(s)}(z)$ depends on the functions
$\rho^{(s)}(z)$ according to (\ref{V_expan}) and the latter depend on
the functions $\phi_n^{(s^\prime)}(z)$ with $s^\prime=0,1,\ldots,s$
according to (\ref{n_s}), and some of these functions
$\phi_n^{(s^\prime)}(z)$ have not yet been computed, even at $s=1$.
We therefore must consider equations (\ref{V_expan}),(\ref{n_s}) and
(\ref{phi}) as a coupled set of equations, which must be solved
self-consistently.

To compute the matrix elements appearing in the right-hand sides of
(\ref{phi}), we use the following procedure. We start with $s=1$ and
define a (yet unknown) quantity 
\begin{equation}
\label{Delta_1}
\Delta_{kn}^{(1)} \equiv \langle \phi_k^{(0)} \vert U_{\rm eff}^{(1)} \vert \phi_n^{(0)} \rangle \ .
\end{equation}
\noindent
In particular, we have $\epsilon_n^{(1)} = \Delta_{nn}^{(1)}$.  By
substituting the expansion (\ref{phi_1}) for $\vert \phi_n^{(1)}
\rangle$ into (\ref{n_1}) and using (\ref{W_expan}) for $W_n^{(1)}$,
we can express $\rho_e^{(1)}(z)$ in terms $\Delta_{kn}^{(1)}$ as
follows:
\begin{align}
\label{rho_1}
\rho_e^{(1)}(z) = & e \sum_{n=1}^{n_{\rm F}} \left\{ w\left [
    \sum_{k=1}^{n_{\rm F}}\Delta_{kk}^{(1)} - n_{\rm
      F}\Delta_{nn}^{(1)}\right ]
  \left(\phi_n^{(0)}(z) \right)^2 + \right. \nonumber \\
& \left. 2 W_n^{(0)} \phi_n^{(0)}(z) \sum_{k\neq
    n}\frac{\Delta_{kn}^{(1)}} {\epsilon_n^{(0)} - \epsilon_k^{(0)}}
  \phi_k^{(0)}(z)\right \} \ .
\end{align}
\noindent
We then substitute (\ref{rho_1}) into (\ref{V_expan}) (in which we
must specialize to the case $s = 1$) and compute the expectation value
of $U_{\rm eff}^{(1)}$ between the unperturbed states $\langle
\phi_n^{(0)} \vert$ and $\vert \phi_n^{(0)} \rangle$ to obtain the
following set of linear equations:
\begin{align}
\Delta_{nm}^{(1)} + & \sum_{k=1}^{n_{\rm F}} \sum_{l\neq k} L_{nm}^{kl}
\Delta_{lk}^{(1)} + w \sum_{k=1}^{n_{\rm F}} \left (\sum_{l=1}^{n_{\rm
      F}}\Delta_{ll}^{(1)} -  n_{\rm F} \Delta_{kk}^{(1)}\right ) \nonumber \\
& \times \langle \phi_n^{(0)} \vert G_{kk}^{(0)}\vert \phi^{(0)}_m \rangle =
R_{nm}^{(1)} \ .
\label{system_1}
\end{align}
\noindent
This set must be solved with respect to the unknown quantities
$\Delta_{nm}^{(1)}$. The matrix $L$ is defined by the following
relations, which contain only the known quantities:
\begin{equation}
\label{Lmnkl}
L_{nm}^{kl} = 2W_k^{(0)}\frac{\langle \phi_n^{(0)} \vert G_{kl}^{(0)} \vert
  \phi_m^{(0)} \rangle}{\epsilon_k^{(0)} - \epsilon_l^{(0)} },
\end{equation}
\begin{align}
\label{Gkl}
G_{kl}^{(0)}(z) & = 2\pi e^2 \int_{-\infty}^{\infty}\vert z -\xi \vert
\phi_k^{(0)}(\xi) \phi_l^{(0)}(\xi) {\rm d}\xi \nonumber \\
& - U_{\rm xc}^\prime(z) \phi_k^{(0)}(z) \phi_l^{(0)}(z)
\end{align}
\noindent
and the right-hand side of the equation is determined by the following
expression:
\noindent
\begin{equation}
\label{R_1}
R_{nm}^{(1)} = \langle \phi_n^{(0)} \vert V^{(1)} \vert \phi_m^{(0)}
\rangle \ .
\end{equation}
\noindent
The computational complexity of solving (\ref{system_1}) can be
reduced by making the following observation. It follows from
(\ref{DeltaV_1}) and (\ref{R_1}) that $R_{m+2q,m}^{(1)} = 0$ for all
$q = 0, \pm 1, \pm 2, \ldots$ and $m, m + 2q \geq 1$. Using
(\ref{parity_0}) and the relation
\begin{equation}
\label{G_0parity}
G_{kl}^{(0)}(-z) = (-1)^{k-l}G_{kl}^{(0)}(z) \ ,
\end{equation} 
\noindent
we find that the set of equations (\ref{system_1}) splits into two
uncoupled subsystems. The first subsystem is homogeneous and contains
the matrix elements of the form $\Delta_{m+2q,m}^{(1)}$. The only
solution to this subsystem is trivial.  Thus, we have
$\Delta_{m+2q,m}^{(1)} = 0$, which also implies that the last term in
the left-hand side of (\ref{system_1}) vanishes and $\epsilon_n^{(1)}
= \Delta_{nn}^{(1)} = 0$. The other subsystem is of the form
\begin{equation}
\label{system_s=1}
\Delta_{m+2q+1 ,m}^{(1)} + 
\sum_{k=1}^{n_{\rm F}}\sum_p L_{m+2q+1,m}^{k+2p+1,k}
\Delta_{k+2p+1,k}^{(1)} = R_{m+2q+1,m}^{(1)}  
\end{equation} 
\noindent
with a nonzero right-hand side. Eq.~(\ref{system_s=1}) must be solved
numerically with respect to the unknowns $\Delta_{m+2q+1,m}^{(1)}$. In
addition to the simplification outlined above, the following parity
relations hold:
\begin{subequations}
\label{parity_1} 
\begin{align}
\label{parity_1phi}
& \phi_n^{(1)}(-z) = (-1)^{n}\phi_n^{(1)}(z)\ , \\
\label{parity_1n}
& \rho_e^{(1)}(-z) = - \rho_e^{(1)}(z) \ , \\
\label{parity_1V}
& U_{\rm eff}^{(1)}(-z) = -U_{\rm eff}^{(1)}(z) \ .
\end{align}
\end{subequations}
\noindent
After solving (\ref{system_s=1}) numerically, one can find $\vert
\phi_n^{(1)} \rangle$ from (\ref{phi_1}), and also $\rho_e^{(1)}(z)$
and $U_{\rm eff}^{(1)}(z)$ from (\ref{n_1}) and (\ref{V_expan}),
respectively.

Next, we take $s = 2$ and start again with computing the matrix
elements $\Delta_{kn}^{(2)}$, which are defined in a manner similar to
(\ref{Delta_1}):
\begin{equation}
\label{Delta_2}
\Delta_{kn}^{(2)} \equiv \langle \phi_k^{(0)} \vert U_{\rm eff}^{(2)} \vert \phi_n^{(0)} \rangle \ .
\end{equation}
\noindent
Once $\Delta_{kn}^{(2)}$ are found, $\vert \phi_n^{(2)} \rangle$ can
be obtained using (\ref{phi_2}) while the second-order correction to
the energy of the $n$-th level is given by
\begin{equation} 
\label{eps_2}
\epsilon_n^{(2)} = \Delta_{nn}^{(2)} + 
\langle \phi^{(0)}_n\vert U_{\rm eff}^{(1)} \vert \phi_n^{(1)} \rangle
\ .
\end{equation}
Note that the terms $\langle \phi_k^{(0)} \vert U_{\rm eff}^{(1)}
\vert \phi_n^{(1)} \rangle$ and $\langle \phi_n^{(1)} \vert
\phi_n^{(1)}\rangle$ in (\ref{phi_2}) are first-order quantities,
which have already been computed.

After substituting (\ref {phi_2}) and (\ref{W_expan}) into
(\ref{n_2}), we arrive at the expression for $\rho_e^{(2)}(z)$, which
is similar to (\ref{rho_1}) and is omitted here. The only unknown
quantities in this expression are the matrix elements
$\Delta_{kn}^{(2)}$. We then substitute this expression into
(\ref{V_expan}), in which we specialize to the case $s = 2$, and take
the expectation between the unperturbed states. This yields the
following set of equations:
\begin{align}
\Delta_{nm}^{(2)} + & \sum_{k=1}^{n_{\rm F}} \sum_{l\neq k} L_{nm}^{kl}
\Delta_{lk}^{(2)} + w \sum_{k=1}^{n_{\rm F}} \left (\sum_{l=1}^{n_{\rm
      F}}\Delta_{ll}^{(2)} -  n_{\rm F} \Delta_{kk}^{(2)}\right ) \nonumber \\
& \times \langle \phi_n^{(0)} \vert G_{kk}^{(0)}\vert \phi^{(0)}_m \rangle =
R_{nm}^{(2)} \ .
\label{system_2}
\end{align}
Here the unknowns are $\Delta_{nm}^{(2)}$. We see that
(\ref{system_2}) has the same matrix as (\ref{system_1}) but a
different right-hand side, viz,
\begin{align}
\label{R_2}
R_{nm}^{(2)} & =  \sum_{k=1}^{n_{\rm F}} \sum_{l\neq k} L_{mn}^{kl} 
\langle \phi^{(0)}_l \vert U_{\rm eff}^{(1)} \vert \phi_k^{(1)}
\rangle \nonumber \\
& + \sum_{k=1}^{n_{\rm F}}W_k^{(0)} \left[ W_k^{(0)} \langle \phi_k^{(1)} \vert
\phi_k^{(1)} \rangle \langle \phi^{(0)}_n \vert G_{kk}^{(0)} \vert
\phi^{(0)}_m \rangle  \right. \nonumber \\
 & \left. - A_k \langle \phi^{(0)}_n \vert G_{kk}^{(0)} \vert
\phi^{(0)}_m \rangle - \langle \phi^{(0)}_n \vert
G_{kk}^{(2)} \vert \phi^{(0)}_m \rangle \right] \nonumber \\
& + \frac{1}{2e^2}
\langle \phi^{(0)}_n \vert U_{\rm xc}^{\prime\prime} [\rho_e^{(1)}]^2 \vert
\phi^{(0)}_m \rangle \ ,
\end{align}
where
\begin{align}
\label{G2kl}
G_{kk}^{(2)}(z) & = 2\pi e^2\int_{-\infty}^{\infty} \vert z -
\xi \vert [\phi_k^{(1)}(\xi)]^2 d \xi \nonumber \\
&  - U_{\rm xc}^{\prime}(z)[\phi_k^{(1)}(z)]^2 
\end{align}
and
\begin{align}
  A_k & = w \left [ \sum_{l=1}^{n_{\rm F}} \langle \phi^{(0)}_l \vert
    U_{\rm eff}^{(1)} \vert \phi_l^{(1)} \rangle - n_{\rm F} \langle
    \phi^{(0)}_k \vert U_{\rm eff}^{(1)} \vert \phi_k^{(1)} \rangle
  \right ] \ .
\label{A}
\end{align}
All quantities, which enter the definition of $R_{nm}^{(2)}$, are
determined by in the first-order. As follows from (\ref{parity_0phi}),
(\ref{G_0parity}), and (\ref{parity_1}), $R_{m+2q+1,m}^{(2)} = 0$ for
all $q = 0, \pm 1, \pm 2, \ldots$ and $m, m + 2q + 1 \geq 1$. Thus,
unlike in the case $s = 1$, we now have a homogeneous subsystem for
$\Delta_{m+2q+1,m}^{(2)}$, which has only the trivial solution, while
the quantities $\Delta_{m+2q,m}^{(2)}$ satisfy the following
inhomogeneous subsystem:
\begin{align}
\label{system_2inhom}
& \Delta_{m+2q,m}^{(2)} + \sum_{l=1}^{n_{\rm F}}\sum_{p\neq 0}
L_{m+2q,m}^{l+2p,l} \Delta_{l+2p,l}^{(2)} \nonumber \\
& + w \sum_{k=1}^{n_{\rm F}} \left (\sum_{l=1}^{n_{\rm
F}}\Delta_{ll}^{(2)} - n_{\rm F}\Delta_{kk}^{(2)}\right )
\langle\phi^{(0)}_{m+2q}\vert G_{kk}^{(0)} \vert \phi^{(0)}_m \rangle
\nonumber \\ & = R_{m+2q,m}^{(2)} \ .
\end{align}  
\noindent
We then solve (\ref{system_2inhom}) with respect to
$\Delta_{m+2q,m}^{(2)}$ numerically and use the result to compute the
second-order corrections, that is, the quantities $\vert \phi_n^{(2)}
\rangle$, $\rho_e^{(2)}$, $U_{\rm eff}^{(2)}$, $W^{(2)}_n$, and
$\epsilon_n^{(2)}$ by using equations (\ref{phi_2}), (\ref{n_2}),
(\ref{V_expan}), (\ref{W_expan}), and (\ref{eps_2}), respectively.  At
the second order, the following parity relations hold:
\begin{subequations}
\label{parity_2} 
\begin{align}
\label{parity_2phi}
& \phi_n^{(2)}(-z) = (-1)^{n-1}\phi_n^{(2)}(z)\ , \\
\label{parity_2n}
& \rho_e^{(2)}(-z) = \rho_e^{(2)}(z)\ , \\
\label{parity_2V}
& U_{\rm eff}^{(2)}(-z) = U_{\rm eff}^{(2)}(z) \ .
\end{align}
\end{subequations}

Finally, in the case $s = 3$, all quantities in the right-hand side of
(\ref{phi_3}) are known except for
\begin{equation}
\label{Delta_3}
\Delta_{kn}^{(3)} \equiv \langle \phi_k^{(0)} \vert U_{\rm eff}^{(3)} \vert \phi_n^{(0)} \rangle \ .
\end{equation}
\noindent
Again, once $\Delta_{kn}^{(3)}$ are found, $\vert \phi_n^{(3)}
\rangle$ can be obtained from (\ref{phi_3}) while the third-order
corrections to the energies are given by
\begin{align} 
\label{epsilon_3}
  \epsilon_n^{(3)} = & \Delta_{nn}^{(3)} + \langle \phi_n^{(0)}\vert
  U_{\rm eff}^{(1)}\vert \phi_n^{(2)} \rangle
  -  \epsilon_n^{(2)}\langle \phi_n^{(0)} \vert \phi_n^{(1)} \rangle
  \nonumber \\
 & + \langle \phi_n^{(0)} \vert U_{\rm eff}^{(2)} \vert
   \phi_n^{(1)} \rangle \ .
\end{align}
\noindent
It follows from the parity relations (\ref{parity_0phi}),
(\ref{parity_1}), and (\ref{parity_2}) that all terms in the
right-hand side of (\ref{epsilon_3}) vanish except for
$\Delta_{nn}^{(3)}$, so that we have $\epsilon_n^{(3)} =
\Delta_{nn}^{(3)}$. In fact, we will see below that $\epsilon_n^{(3)}
= \Delta_{nn}^{(3)} = 0$. Similarly to the procedure outlined above
for the $s=1$ and $s=2$ cases, we substitute $\phi_n^{(3)}$ from
(\ref{phi_3}) and $W^{(3)}_n$ from (\ref{W_expan}) into (\ref{n_3}),
obtain an expression for $\rho_e^{(3)}(z)$, in which the only unknown
quantities are $\Delta_{kn}^{(3)}$, and substitute the result into
(\ref{V_expan}), where we specialize now to the case $s = 3$. Finally,
we compute the same expectations as before and obtain the set of
equations
\begin{align}
\Delta_{nm}^{(3)} + & \sum_{k=1}^{n_{\rm F}} \sum_{l\neq k} L_{nm}^{kl}
\Delta_{lk}^{(3)} + w \sum_{k=1}^{n_{\rm F}} \left (\sum_{l=1}^{n_{\rm
      F}}\Delta_{ll}^{(3)} -  n_{\rm F} \Delta_{kk}^{(3)}\right ) \nonumber \\
& \times \langle \phi_n^{(0)} \vert G_{kk}^{(0)}\vert \phi^{(0)}_m \rangle =
R_{nm}^{(3)}  
\label{system_3}
\end{align}
with respect to the unknowns $\Delta_{nm}^{(3)}$. The matrix is the
same as before, while the right-hand side is given by the following
relations:
\begin{subequations}
\begin{align}
\label{R_3}
R_{nm}^{(3)} = & -2\sum_{k=1}^{n_{\rm F}}W_k^{(0)} \left[ \sum_{l\neq
    k} D_{kl} \langle \phi^{(0)}_n \vert G_{kl}^{(0)} \vert
  \phi^{(0)}_m \rangle \nonumber \right. \\
& \left. - \langle \phi^{(0)}_n \vert G_{kk}^{(3)} \vert \phi^{(0)}_m
  \rangle \right] - 2\sum_{k=1}^{n_{\rm F}}W_k^{(2)} \langle
\phi^{(0)}_n \vert
G_{kk}^{(1)} \vert \phi^{(0)}_m \rangle \nonumber \\
& + \frac{1}{e^2} \langle \phi^{(0)}_n \vert U_{\rm xc}^{\prime\prime}
\rho_e^{(1)}\rho_e^{(2)} \vert \phi^{(0)}_m
\rangle \nonumber \\
& - \frac{1}{6e^3} \langle \phi^{(0)}_n \vert U_{\rm
  xc}^{\prime\prime\prime} [\rho_e^{(1)}]^3 \vert \phi^{(0)}_m \rangle
\ .
\end{align}
\end{subequations}
\begin{align}
\label{Gkll}
G_{kk}^{(1)}(z) & = 2\pi e^2\int_{-\infty}^{\infty} \vert z -
\xi \vert \phi_k^{(1)}(\xi) \phi_k^{(0)}(\xi)d \xi  \nonumber \\ 
& - U_{\rm xc}^\prime(z) \phi_k^{(1)}(z) \phi_k^{(0)}(z)  \ ,
\end{align}
\begin{align}
\label{G3kl}
G_{kk}^{(3)}(z) & = 2\pi e^2\int_{-\infty}^{\infty} \vert z -
\xi \vert \phi_k^{(1)}(\xi) \phi_k^{(2)}(\xi) d\xi  \nonumber \\
& - U_{\rm xc}^\prime(z)\phi_k^{(1)}(z)\phi_k^{(2)}(z) \ ,
\end{align}
where
\begin{align}
& D_{nk} = \nonumber \\ 
& \frac{\langle \phi_k^{(0)}\vert U_{\rm eff}^{(1)}\vert \phi_n^{(2)} \rangle +
  \langle \phi_k^{(0)} \vert U_{\rm eff}^{(2)} \vert \phi_n^{(1)} \rangle - 
 \epsilon_n^{(2)} \langle \phi_k^{(0)} \vert \phi_n^{(1)} \rangle}{\epsilon_n^{(0)} -
 \epsilon_k^{(0)}} \ .
\label{Cbar}
\end{align}
\noindent
Similarly to the case $s = 1$, we have $R_{m+2q,m}^{(3)} = 0$ for all
$q = 0, \pm 1, \pm 2, \ldots$ and $m, m + 2q \geq 1$. This follows
from the parity relations (\ref{parity_0phi}), (\ref{G_0parity}),
(\ref{parity_1}), and (\ref{parity_2}). Consequently, we have
$\Delta_{m+2q,m}^{(3)} = 0$ and $\epsilon_n^{(3)} = \Delta_{nn}^{(3)}
= 0$. The matrix elements of the form $\Delta_{m+2q+1,m}^{(3)}$ are
determined from the following subset:
\begin{equation}
\label{system_1_3inhom}
\Delta_{m+2q+1 ,m}^{(3)} + 
\sum_{k=1}^{n_{\rm F}}\sum_p L_{m+2q+1,m}^{k+2p+1,k}
\Delta_{k+2p+1,k}^{(3)} = R_{m+2q+1,m}^{(3)} \ . 
\end{equation} 
\noindent
As in the $s = 1$ case, the following parity relations hold:
\begin{subequations}
\label{parity_3} 
\begin{align}
\label{parity_3phi}
& \phi_n^{(3)}(-z) = (-1)^{n}\phi_n^{(3)}(z)\ , \\
\label{parity_3n}
& \rho_e^{(3)}(-z) = - \rho_e^{(3)}(z) \ , \\
\label{parity_3V}
& U_{\rm eff}^{(3)}(-z) = -U_{\rm eff}^{(3)}(z) \ .
\end{align}
\end{subequations}
\noindent
By solving (\ref{system_1_3inhom}) we can find, in particular, the
expansion coefficient $\rho_e^{(3)}(z)$ for the negative charge
density from (\ref{n_3}) using (\ref{phi_3}) and (\ref{W_expan}) at $s
= 3$. The expansion coefficients $\alpha_1$ and $\alpha_3$ in
(\ref{exp_1}) are then obtained from
\noindent
\begin{equation}
\label{dipole_coef}
\alpha_{s} = -\frac{4\pi}{h{\mathscr E}_{\rm at}} \int_{-\infty}^{\infty} z \rho_e^{(s)}(z) dz \ .
\end{equation}
\noindent
It can be seen from (\ref{parity_2n}) that $\alpha_2 = 0$.

\section{Results}
\label{sec:res}

We have performed computations for varying film width and different
boundary conditions. We have found that, in the case of R- and B-type
boundary conditions, the effect of including the exchange-correlation
potential is relatively minor. However, in the case of the F-type
boundary condition, the exchange-correlation potential must be
included in order to stabilize the conduction electrons.  For the
purposes of a fair comparison, we show the results of R- and B-type
simulations without the exchange-correlation potential, except in
Fig.~3 below, where the results with and without the
exchange-correlation potential are compared. It will be shown that the
R- and B-type boundary conditions produce a negative nonlinear
correction to the polarizability and saturation effects, in agreement
with Refs.~\onlinecite{hache_86_1,rautian_97_1,panasyuk_08_1}.
However, the F-type boundary condition results in a positive nonlinear
correction whose magnitude is a few hundred times larger than that in
the case of R- or B-type boundary conditions. In all cases, the
emergence of macroscopic (bulk) behavior becomes evident in relatively
wide films.

The eigenproblem~(\ref{ansatz_z}) was solved algebraically by
discretizing the differential equation in the interval $-z_{\rm max}/2
\leq z \leq z_{\rm max}/2$. Here $z_{\rm max} = h + 12a = (M+12)a$ for
the F-type boundary condition, $z_{\rm max} = h + 2\Delta_B$ for the
B-type boundary condition ($\Delta_B$ is defined in (\ref{Delta_B})),
and $z_{\rm max} = h$ for the R-type boundary condition. Recall that
the jellium (the physical slab) is contained in the region $-h/2 < z <
h/2$, where $h=Ma$. Central differences with $20$ discrete points per
the lattice unit $a$ have been used and convergence was verified by
doubling this number.

\begin{figure}
\psfig{file=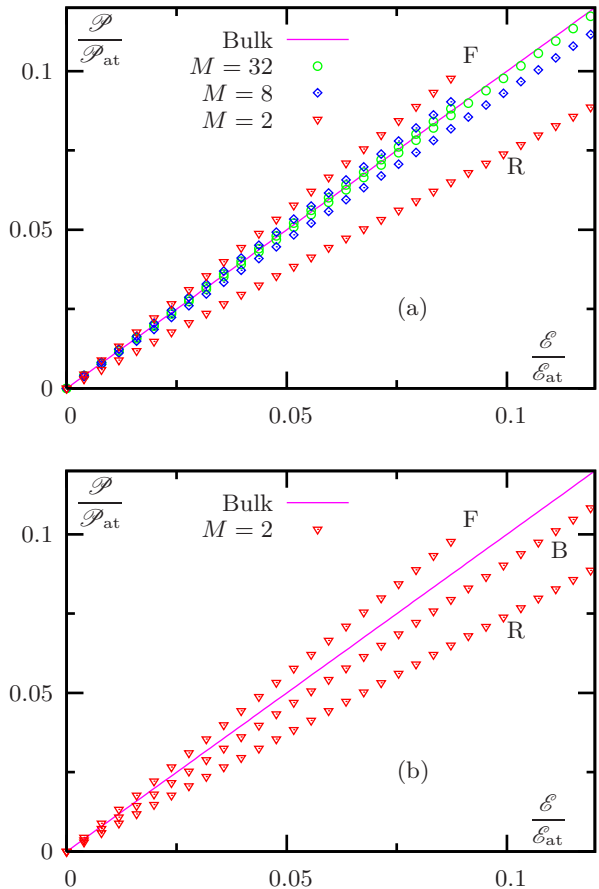,width=8.2cm,bbllx=70bp,bblly=430bp,bburx=300bp,bbury=800bp,clip=t}
\caption{The dipole moment density per unit area,
  ${\mathscr P}$, as a function of the applied field, ${\mathscr E}$,
  for different types of boundary conditions and for films consisting
  of different numbers of atomic layers, $M$. The continuous line
  corresponds to the ``bulk'' (macroscopic) result and the centered
  symbols represent the results obtained from the DFT. In panel (a),
  three different values of $M$ are used, as labeled. All data points
  above the continuous line correspond to F-type boundary conditions
  and the points below the line correspond to R-type boundary
  conditions; B-type boundary conditions are not used in this panel.
  In panel (b), only $M=2$ is used. The two sets of centered symbols
  below the continuous line in (b) correspond to R- and B-type
  boundary conditions, as labeled. In both panels, the data points for
  F-type boundary conditions (above the continuous line) terminate at
  ${\mathscr E}/{\mathscr E}_{\rm at} = 0.87$; computations for larger
  values of ${\mathscr E}$ (with F-type boundary conditions) are
  affected by the numerical instability discussed in the text.}
\end{figure}

In Fig.~1, we plot ${\mathscr P}$ as a function of the applied field,
${\mathscr E}$, for different widths of the slab and for different
types of boundary conditions. In the case of the R- or B-type boundary
conditions, the system is stable independent of the strength of the
applied field. In the case of F-type boundary condition, we have
observed an instability of the charge density for ${\mathscr E}\gtrsim
0.1 {\mathscr E}_{\rm at}$. This instability can be explained by the
effect of tunneling, which can result in significant charge
accumulation in a biased potential over long periods of time, or after
many DFT iterations. The data points affected by this instability are
not displayed in Fig.~1. It can be seen that the F-type boundary
condition tends to increase ${\mathscr P}$ (compared to the
macroscopic limit) while the B- or R-type boundary conditions tend to
decrease ${\mathscr P}$. For all types of boundary conditions,
deviations from the macroscopic result are significant when $M=2$ and
$M=8$ but small when $M=32$. In the case $M=2$ (Fig.~1b), the B-type
curve lies above the R-type curve; a similar result was obtained for
other values of $M$ (data not shown).  Physically, the behavior
illustrated in Fig.~1 can be understood as the result of electron
spillover~\cite{pustovit_06_1,pustovit_06_2} (in the case of F-type
boundary condition) or as the combined action of the finite charge
density of the jellium and of the uncertainty principle (in the case
of B- or R-type boundary conditions).

Although the deviation of ${\mathscr P}$ from the macroscopic result
is obvious in Fig.~1, all curves shown in this figure appear to be
linear. To visualize the deviations from linearity, we have computed
the nonlinear contribution to the dipole moment density according to
\begin{equation}
{\mathscr P}_{\rm nonl} = {\mathscr P} - \frac{h\alpha_1}{4\pi}{\mathscr
  E} \ . 
\end{equation}
\noindent
The result is plotted in Fig.~2. It can be seen that the correction is
negative for R- and B-type boundary conditions. For the F-type
boundary condition, the correction is positive and about $200$ times
larger in magnitude.  An interesting effect can be seen in Fig.~2b.
Namely, the B-type boundary condition produces a nonlinear correction
of a larger magnitude compared to the R-type boundary condition.  This
is somewhat unexpected since the data of Fig.~1 suggest that the
B-type curve is closer to the macroscopic asymptote.  Moreover, we
have discovered a non-monotonic dependence of the expansion
coefficient $\alpha_3$ in (\ref{exp_1}) on the displacement parameter
$\Delta$, as is discussed below.

\begin{figure}
\psfig{file=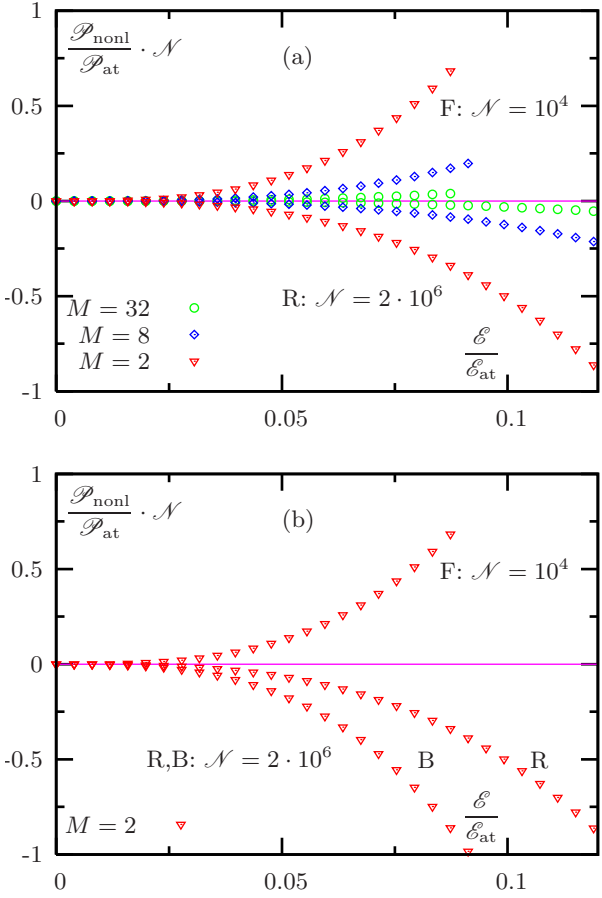,width=8.2cm,bbllx=70bp,bblly=430bp,bburx=300bp,bbury=800bp,clip=t}
\caption{${\mathscr P}_{\rm nonl}$ as a function
  of the applied field, ${\mathscr E}$. The same convention for
  encoding the different types of boundary conditions and the
  different numbers of atomic layers as in Fig.~1 is used.
  Additionally, different numerical normalization factors ${\mathscr
    N}$ have been used for different types of boundary conditions, as
  labeled.}
\end{figure}

Recall that the B-type boundary condition involves a displacement of
the rigid potential wall from the surface of the metal by the distance
$\Delta=\Delta_{\rm B} \approx 0.38\ell$.  We can, however, view
$\Delta$ as a free parameter. In the case $\Delta=0$, we recover the
R-type boundary condition, in the case $\Delta=\infty$ -- the F-type
boundary condition, and $\Delta=\Delta_{\rm B}$ corresponds to
Bardeen's model. In Fig.~3, we plot the expansion coefficient
$\alpha_3$ as a function of $\Delta$ for $M=2$. In this figure, we
show the data obtained both with and without the exchange-correlation
potential. We find that the surprising non-monotonic dependence is
observed in both cases. For larger values of $\Delta$, the red curve
in Fig.~3 (with exchange-correlation potential included) rapidly grows
and saturates at the level of $\alpha_3\approx 0.1$ for $\Delta
\gtrsim 8 \Delta_{\rm B}$ (data not shown). The latter result exactly
corresponds to the one obtained with the F-type boundary condition.
Note that $\alpha_1$ is almost independent of $\Delta$. Qualitatively
the same results have been obtained for $M=8$ and $M=32$.

\begin{figure}
\psfig{file=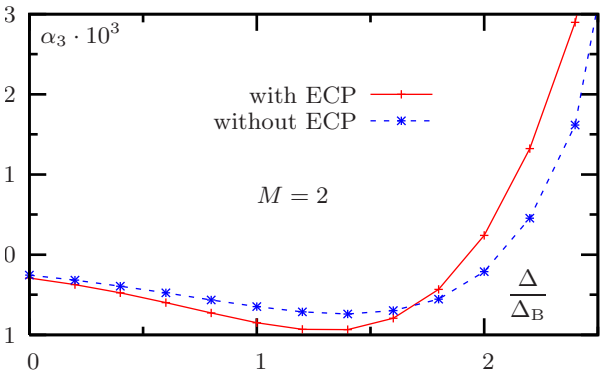,width=8.2cm,bbllx=70bp,bblly=630bp,bburx=300bp,bbury=800bp,clip=t}
\caption{Dependence of $\alpha_3$ on $\Delta$ for
  $M=2$, with and without accounting for the exchange-correlation
  potential (ECP).}
\end{figure}

Next, in Fig.~4, we plot ${\mathscr P}$ for the B-type boundary
condition (with $\Delta=\Delta_{\rm B}$) in a very large interval of
${\mathscr E}$, up to ${\mathscr E}/{\mathscr E}_{\rm at} = 10^3$.  Of
course, an applied static electric field of this magnitude is not
achievable in practice.  However, the situation can be more
experimentally favorable in the case of {\em quasistatic} fields.
Although we do not consider this case directly, it is known that the
internal field enhancement factor due to plasmon resonances is of the
order of $\omega_p/\gamma$, where $\omega_p$ is the plasma frequency
and $\gamma$ is the Drude relaxation constant. This factor can be as
large as $\sim 500$ in the case of silver, and it enters the nonlinear
correction to the polarizability in the fourth
power~\cite{drachev_04_2,govyadinov_11_1}. Note that the quasistatic
approximation (known in the context of DFT as the adiabatic
approximation) is applicable as long as $h\omega/c \ll 1$, which
easily holds even in the visible spectral range.  Also, the electric
field intensity in very short laser pulses can be of the order of or
higher than the atomic field, and the related physics has attracted
considerable recent attention~\cite{durach_10_1}. 

\begin{figure}
\psfig{file=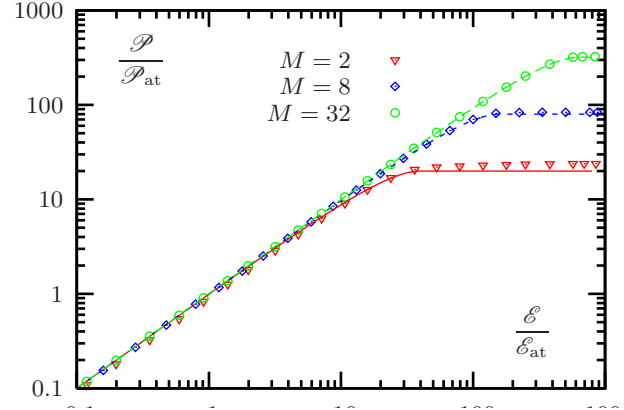,width=8.2cm,bbllx=70bp,bblly=630bp,bburx=300bp,bbury=800bp,clip=t}
\caption{${\mathscr P}$ as a function of the
  applied field, ${\mathscr E}$, for B-type boundary condition and for
  different numbers of atomic layers, $M$. Centered symbols correspond
  to DFT results and continuous lines to the analytical expression
  (\ref{P_class}).}
\end{figure}

In Fig.~4, we also compare the DFT calculations with the expression
\begin{equation}
\label{P_class}
{\mathscr P} = \left\{
\begin{array}{ll}
{\displaystyle \frac{h{\mathscr E}}{4\pi}} \left[1 - {\displaystyle \frac{\ell}{8\pi
      h} \left\vert \frac{\mathscr E}{\mathscr  E}_{\rm at}
    \right\vert} \right] \ , &  {\rm if} 
\left\vert {\displaystyle \frac{\mathscr E}{\mathscr E}_{\rm at}} \right\vert \leq
{\displaystyle \frac{4\pi h}{\ell}} \ ,  \\
{\displaystyle \frac{h^2 {\mathscr E}_{\rm at}}{2\ell}} \ , & \ \ {\rm otherwise} \ ,
\end{array}
\right.
\end{equation}
\noindent
which was derived in Ref.~\onlinecite{panasyuk_08_1} using purely
classical arguments.  It can be seen that (\ref{P_class}) is
surprisingly accurate for $M=8$ and especially for $M=32$.  This may
seem unexpected because (\ref{P_class}) contains a nonanalyticity of
the form ${\mathscr E} \vert {\mathscr E} \vert$, while the expansion
(\ref{exp_1}) represents a real analytic function. This discrepancy is
resolved by noting that (\ref{exp_1}) has a finite radius of
convergence and that an expansion of this type can not, in principle,
capture the saturation phenomena illustrated in Fig.~3. On the other
hand, numerical DFT calculations can be carried out whether or not
(\ref{exp_1}) converges.

Finally, we investigate the dependence of the coefficients $\alpha_1$,
$\alpha_3$ on the number of atomic layers $M$ for all three types of
boundary conditions. The results are shown in Fig.~5. It can be seen
that, in all cases, the dependence is monotonic. Comparing the results
for R- and B-type boundary conditions, we reconfirm the trend that has
been already noted, namely, that B-type boundary conditions produce a
smaller finite-size correction to $\alpha_1$ (compared to R-type
boundary conditions) but a larger nonlinear response.

\begin{figure}
\psfig{file=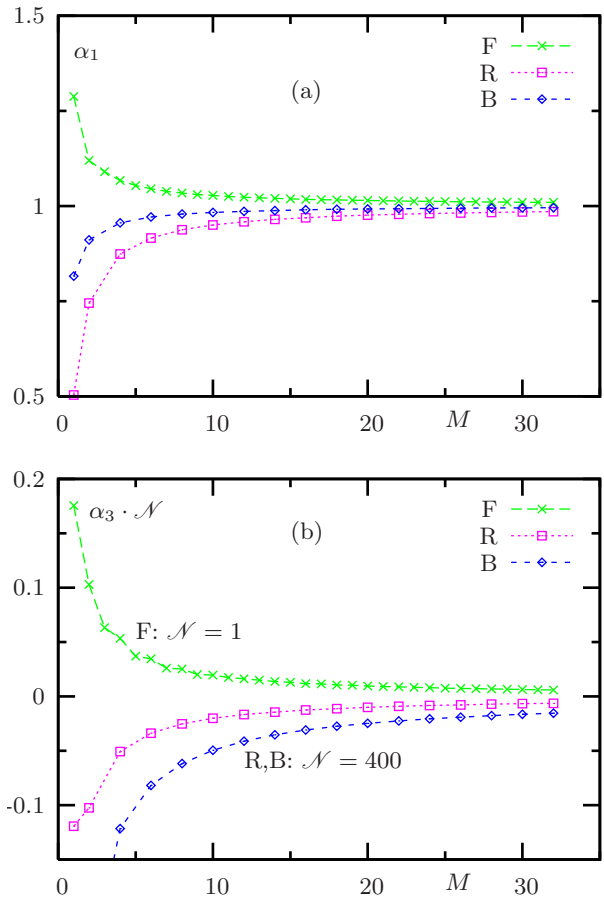,width=8.2cm,bbllx=70bp,bblly=430bp,bburx=300bp,bbury=800bp,clip=t}
\caption{Coefficients $\alpha_1$ (a) and
  $\alpha_3$ (b) as functions of the number of atomic layers, $M$, for
  three different types of boundary conditions. Different numerical
  normalization factors ${\mathscr N}$ have been used for different
  types of boundary conditions, as labeled.}
\end{figure}

\section{Summary and discussion}
\label{sec:summ}

We have studied theoretically and numerically polarization of a thin
silver film under perpendicularly applied low-frequency external
electric field. Three different boundary conditions have been applied
at the film surface. It was shown that the sign and magnitude of the
nonlinear correction to the film polarizability depends dramatically
on the type of boundary condition used. Since all theories involved
contain approximations, only comparison with experiment can determine
which boundary condition is physically correct.

An obvious shortcoming of the calculations reported herein is that
they are carried out for static fields. However, the results can be
extended to finite frequencies, as long as relaxation and resonance
phenomena are not taken into consideration, that is, if the frequency
is far below the lowest plasmon resonance of the system. In practice,
this means that the theory can be applied up to THz frequencies. Thus
our results are amenable to experimental verification. A possible
experimental test could be a measurement of the sign of the real part
of the nonlinear susceptibility $\chi^{(3)}$ for a suspension of
silver nanodisks at the excitation frequency $\sim 10{\rm GHz}$.  The
disk thickness should be much smaller than the skin depth, $\delta
\approx 0.2\mu{\rm m}$ in this example. Previous experimental
measurements of $\chi^{(3)}$ were largely confined to the optical and
near-IR spectral
regions~\cite{uchida_94_1,petrov_96_1,danilova_96_1,torres-torres_07_1},
where the sign of ${\rm Re}[\chi^{(3)}]$ can depend on frequency due
to the effects of plasmon resonances.

It seems possible to further extend our theory to optical frequencies
by utilizing the quasistatic approximation, which is known as the
adiabatic approximation in the context of DFT~\cite{runge_84_1}. In
this approximation, all potentials are computed using instantaneous
values of the density, for example, by writing for the Hartree
interaction potential $U_{\rm H}[\rho_e]({\bf r},t) = U_{\rm H}
[\rho_e({\bf r},t)]$, and similarly for other functionals. This
corresponds to neglecting the effects of retardation in the
electromagnetic interaction and is adequate as long as $h\omega/c \ll
1$. Note that plasmon resonances and relaxation phenomena can be taken
into consideration within quasistatics. However, time-dependent DFT is
still relatively unexplored, although some promising results have been
obtained~\cite{vasiliev_02_1}.

This work was supported by the NSF under the grant DMR0425780. One of
the authors (GYP) is supported by the National Research Council Senior
Associateship Award at the Air Force Research Laboratory.

\bibliographystyle{apsrev} 
\bibliography{abbrev,master,book,local}

\end{document}